\documentclass[epj,final]{svjour}
\usepackage{graphics}
\input{epsf}
\begin{document}
\title{Statistical Properties of Eigenvalues 
for an Operating Quantum Computer with Static Imperfections} 
\author{Giuliano Benenti\inst{1}, Giulio Casati\inst{1,2},
Simone Montangero\inst{1}, and Dima L. Shepelyansky\inst{3}}
\institute{International Center for the Study of Dynamical
Systems, Universit\`a degli Studi dell'Insubria and
Istituto Nazionale per la Fisica della Materia,
Unit\`a di Como, Via Valleggio 11, 22100 Como, Italy
\and
Istituto Nazionale di Fisica Nucleare,
Sezione di Milano, Via Celoria 16, 20133 Milano, Italy
\and
Laboratoire de Physique Quantique, UMR 5626 du CNRS,
Universit\'e Paul Sabatier, 31062 Toulouse Cedex 4, France}
\titlerunning{Eigenvalues of an operating quantum computer}
\authorrunning{G. Benenti, G. Casati, S.Montangero, and
D.L. Shepelyansky}
\date{Received: June 19, 2002}
\abstract{We investigate the transition to quantum chaos,
induced by static imperfections, for an operating quantum 
computer that simulates efficiently a dynamical quantum system, 
the sawtooth map.
For the different dynamical regimes of the map, we discuss 
the quantum chaos border induced by static imperfections 
by analyzing the 
statistical properties of the quantum computer 
eigenvalues. For small imperfection strengths the level spacing statistics
is close to the case of quasi-integrable systems while above
the border it is described by the random matrix theory. 
We have found that the border drops exponentially with the number of qubits,
both in the ergodic and quasi-integrable dynamical
regimes of the map characterized by a complex phase space 
structure. On the contrary, the regime with integrable map dynamics
remains more stable against static imperfections since in this case the
border drops only algebraically with the number of qubits. 
\PACS{
{03.67.Lx}{Quantum computation} \and
{05.45.Mt}{Semiclassical chaos ("quantum chaos")} \and
{24.10.Cn}{Many-body theory}
}
}

\maketitle

\section{Introduction}

Quantum computers, if constructed, would be capable of 
solving some computational problems much more efficiently 
than classical computers \cite{chuang}. 
Shor \cite{shor} constructed a quantum algorithm which performs
integer factorization exponentially faster than any
known classical algorithm. It was also shown by Grover
\cite{grover} that the search of an item in an unstructured list can be
done with a square root speedup over any classical algorithm.
These results motivated a great body of experimental proposals for the
construction of a realistic quantum computer (see \cite{chuang} and
references therein).

While the technological challenge to develop scalable, fault tolerant
quantum processors is highly demanding, it is well appreciated
that decoherence, due to the coupling with the environment, will be the
ultimate obstacle to the realization of such devices.
In addition, even in the ideal case in which the quantum computer is
isolated from the external world, a proper operability of the computer
is not guaranteed: unavoidable internal  static imperfections in the quantum
computer hardware represent another source of errors.
For example, the energy spacing between the two states of each qubit can
fluctuate, e.g. due to magnetic field inhomogeneities in
nuclear magnetic resonance quantum processors \cite{chuang}.
Moreover, since qubit interactions are required to operate
two-qubit gates and generate entangled states, unwanted residual
interactions will appear.
For example, when the inter-qubit coupling is switched off, 
e.g. via a potential barrier created by a point contact gate
in the quantum dots proposal \cite{loss}, some unavoidable
residual interactions still remain.
Therefore the quantum computer hardware can be modeled as a 
qubit lattice  and one has to consider it as a quantum 
many-qubit (-body) interacting system \cite{GS}.
Many-body quantum  systems have been widely investigated in the field of
quantum chaos \cite{bohigas,guhr} and it
is now well known that residual interactions can lead to quantum
chaos characterized by ergodicity of the eigenstates and level
spacing statistics described by random matrix theory 
\cite{aberg,jacquod,JLP,flambaum}.
In the regime of quantum chaos 
the wave functions and energy spectra become
so complicated that a statistical description should be
applied to them.
Thus it is important to study the stability of quantum information
processing in the presence of realistic models of quantum
computer hardware imperfections and in concrete examples of 
quantum algorithms.
The stability of the static   quantum 
computer hardware was studied in \cite{GS} and it was shown 
that residual inter-qubit interactions can destroy 
a generic state stored in the quantum computer 
if the amplitude of static imperfections is above the quantum chaos border.
For a static quantum hardware this border drops linearly
with the number of qubits while the average energy 
level spacing drops exponentially. Above the quantum chaos border
an exponential number of ideal multi-qubit states becomes
mixed after a finite chaotic time scale.  
The implications of such imperfections for the Grover
algorithm and quantum Fourier transform 
were analyzed in Ref. \cite{braun}.

In this paper, we address the question of the transition to 
chaos for an {\it operating} quantum computer, which is 
running an efficient quantum algorithm simulating 
the dynamical evolution of the so-called quantum 
sawtooth map \cite{simone1}. 
The sawtooth map is a paradigm of classical \cite{sawc} 
and quantum \cite{sawq} chaos, and exhibits a variety of 
different behaviors, from anomalous diffusion to quantum 
ergodicity and dynamical localization. We note that in the 
above quantum algorithm the number of redundant qubits can be 
reduced to zero and complex dynamics can 
be investigated already with less than 10 qubits.
It is therefore important to study the stability of 
such algorithm, in view of its possible implementation 
in the first generation of quantum processors 
operating with a small number of qubits \cite{qftexp,shorexp}.

Since spectral statistics is the most valuable tool to detect
the transition to quantum chaos \cite{bohigas,guhr}, we study the 
statistical properties of the eigenvalues of the quantum 
sawtooth map, simulating different dynamical regimes 
(ergodic, quasi-integrable, and integrable), in the presence 
of static imperfections in the quantum computer hardware. 
We will show that the threshold for the transition to chaos 
induced by static imperfections drops exponentially with the 
number of qubits both in the ergodic and in the quasi-integrable 
regime, while in the integrable regime the chaos border 
drops only polynomially with the number of qubits. 
We note that this paper complements Ref. \cite{simone2}, devoted 
to the study of the eigenvectors of the model in the 
ergodic regime.   

The paper is organized as follows. In Section 2 we discuss 
the quantum algorithm simulating the quantum sawtooth map model 
and its realization in the presence of hardware static 
imperfections.
The effect of these imperfections on the spectral statistics 
of the model is described in Sections 3,4, and 5, 
for the ergodic, quasi-integrable, and integrable regime,
respectively. Our conclusions are presented in Section 6. 

\section{The model and the quantum algorithm}

The quantum sawtooth map is the quantized version of the classical sawtooth
map, which is given by
\begin{equation}
\overline{n}={n}+k(\theta-\pi),
\quad
\overline{\theta}=\theta+T\overline{n},
\label{clmap}
\end{equation}
where $(n,\theta)$ are conjugated action-angle variables
($0\le \theta <2\pi$), and the bars denote the 
variables after one map iteration. 
Introducing the rescaled momentum 
variable $p=Tn$, one can see that the classical dynamics depends only on
the single parameter $K=kT$. As it is known, the classical motion is 
stable for $-4<K<0$ and completely chaotic for $K<-4$ and $K>0$
\cite{sawc}.

The quantum evolution for one map iteration is described
by a unitary operator $\hat{U}_0$ (Floquet operator) 
acting on the wave function $\psi$:
\begin{equation}
\overline{\psi}=\hat{U}_0 \psi =
e^{-iT\hat{n}^2/2}
e^{ik(\hat{\theta}-\pi)^2/2}\psi,  
\label{qumap}
\end{equation}
where $\hat{n}=-i\partial/\partial\theta$ 
and $\psi(\theta + 2\pi) = \psi(\theta)$ (we set $\hbar=1$).
The classical limit corresponds to $k\to \infty$, 
$T\to 0$, and $K=kT=\hbox{const}$. 

In this paper, we study the quantum sawtooth map (\ref{qumap}) closed on 
the torus $-\pi \le p <\pi$. Therefore the classical limit is obtained by 
increasing the number of qubits $n_q=\log_2 N$ 
($N$ is the total number of levels), 
with $T=2\pi/N$ ($k=K/T$, $-N/2 \le n < N/2$). We consider 
the ergodic, quasi-integrable and integrable regimes. To this 
end, we take $K=\sqrt{2}$, $K=-0.1$, and $K=-1$, respectively.  

The quantum algorithm introduced in \cite{simone1} simulates 
efficiently the quantum dynamics (\ref{qumap}) using 
a register of $n_q$ qubits. 
It is based on the forward/backward quantum Fourier transform 
\cite{qft} between the $\theta$ and $n$ representations
and has some elements of the quantum algorithm for kicked rotator
\cite{GS1}. 
Such an approach is rather convenient since
the Floquet operator  $\hat{U}_0$ is the product of two operators 
$\hat U_k = e^{ik(\hat{\theta}-\pi)^2/2}$ and $\hat U_T = e^{-iT\hat{n}^2/2}$:
the first one is diagonal in the $\hat \theta$ representation, 
the latter in the 
$\hat n $ representation. 
Thus the quantum algorithm for one map iteration requires the following steps:
\\
{\sc I.}  The unitary operator $\hat U_k$ is decomposed 
in $n_q^2$ two-qubit gates 
\begin{equation}
e^{\imath k(\theta -\pi)^2/2} = \prod_{i,j} 
e^{\imath 2 \pi^2 k( \alpha_i 2^{-i} - 
\frac{1}{2n_q})
(\alpha_j 2^{-j} - 
\frac{1}{2n_q})},
\label{dec}
\end{equation}
where $\theta = 2\pi\sum \alpha_i 2^{-i}$, with $ \alpha_i \in \{ 0,1 \} $. 
Each two-qubit gate can be written in the 
$\{|00 \rangle,|01 \rangle,|10 \rangle,|11 \rangle\}$ basis as  
$\exp(i k \pi^2 D)$, where D is a diagonal matrix with elements 
\begin{eqnarray} 
\{
\frac{1}{2 n_q^2}, 
-\frac{1}{n_q}\left(\frac{1}{2^j}-\frac{1}{2n_q}\right) ,
-\frac{1}{n_q}\left(\frac{1}{2^i}-\frac{1}{2n_q}\right) ,\nonumber\\
2\left(\frac{1}{2^i}-\frac{1}{2n_q}\right)
\left(\frac{1}{2^j}-\frac{1}{2n_q}\right)
\}. 
\end{eqnarray} 
{\sc II.} The change from the $\theta$ to the $n$ representation 
is obtained by means 
of the quantum Fourier transform, which requires $n_q$ Hadamard gates and   
$n_q (n_q -1)/2$ controlled-phase shift gates \cite{qft}.\\
{\sc III.} In the new representation the operator $\hat U_T$ has essentially the 
same form as $\hat U_k$ in step {\sc I} and therefore it can be decomposed in 
$n_q^2$ gates similarly to equation (\ref{dec}). \\
{\sc IV.}  We go back to the initial $\theta$ representation via inverse 
quantum Fourier transform.\\
On the whole the algorithm requires $3 n_q^2 +n_q$ gates per map iteration.   
Therefore it is exponentially efficient with respect to any known classical 
algorithm. Indeed the most efficient way to simulate the quantum dynamics 
(\ref{qumap}) on a classical computer 
is based on forward/backward fast Fourier transform and requires 
$O(n_q 2^{n_q})$ operations. We stress that this quantum algorithm does not need any
extra work space qubit. This is due to the fact that for the quantum sawtooth map 
the kick operator $\hat U_k$ has the same quadratic 
form as the free rotation operator 
$\hat U_T$.

We model the quantum computer hardware as an linear array of qubits (spin 
halves) with static imperfections, i.e. fluctuations in the individual 
qubit energies and residual short-range inter-qubit couplings \cite{GS}. 
The model is described by the following many-body Hamiltonian:  
\begin{equation}
\hat{H}_{\hbox{s}}=\sum_i (\Delta_0+\delta_i)\hat{\sigma}_i^z +
\sum_{i<j}J_{ij}\hat{\sigma}_i^x\hat{\sigma}_j^x,
\label{imperf}
\end{equation}
where the $\hat{\sigma}_i$ are the Pauli matrices for the qubit $i$,
and $\Delta_0$ is the average level spacing for one qubit. 
The second sum in (\ref{imperf}) runs over nearest-neighbor qubit 
pairs, zero boundary conditions are applied, 
and $\delta_i$, $J_{ij}$ are randomly and uniformly distributed
in the intervals $[-\delta/2,\delta/2]$ and $[-J,J]$, respectively.
We model the implementation of the above described algorithm as 
a sequence of instantaneous and perfect one- and two-qubit gates, separated by 
a time interval $\tau_g$,
during which the hardware Hamiltonian (\ref{imperf}) gives unwanted
phase rotations and qubit couplings. Therefore we study numerically the many-body Hamiltonian 
\begin{equation}
\hat H( \tau ) = \hat H_{\hbox{s}} + \hat H_{\hbox{g}}( \tau ), 
\end{equation}
where 
\begin{equation}
\hat H_{\hbox{g}}(\tau ) = \sum_k \delta (\tau  - k \tau_g ) \hat h_k.
\end{equation}
Here $\hat h_k$ realizes  the $k$-th elementary gate according to the  
sequence prescribed by the above algorithm.    
We assume that the average phase accumulation given by 
$\Delta_0$ is eliminated, e.g. by means of refocusing 
techniques \cite{chuang2}.  

Since the evolution operator (\ref{qumap}) remains periodic 
in the presence of static imperfections, 
$\hat{U}_{\epsilon,\rho}(\tau+T)=\hat{U}_{\epsilon,\rho}(\tau)$ 
($\epsilon\equiv \delta \tau_g$,  
$\rho\equiv J \tau_g$ 
rescaled imperfection strengths), 
all information about the system dynamics is included in 
the quasi-energy eigenvalues 
$\lambda_\alpha^{(\epsilon,\rho)}$  
and eigenstates 
$\phi_\alpha^{(\epsilon,\rho)}$ 
of the perturbed Floquet operator: 
\begin{equation} 
\hat{U}_{\epsilon,\rho}(T)  
\phi_\alpha^{(\epsilon,\rho)}=\exp(i \lambda_\alpha^{(\epsilon,\rho)}) 
\phi_\alpha^{(\epsilon,\rho)}.  
\label{floquet} 
\end{equation} 
We study numerically the spectral statistics of the Floquet operator
(\ref{floquet}) for a quantum computer running the quantum sawtooth map 
algorithm in the presence of static imperfections described by (\ref{imperf}). 
We construct numerically the Floquet operator in the momentum representation 
(i.e. the quantum register states basis) using the fact that 
the one step map evolution 
(including static imperfections) of each quantum register state 
gives a column in the matrix representation of this operator.

We consider $4 \le n_q \le 12$ qubits. In order to reduce statistical 
fluctuations, data are averaged over $ 3 \le N_D \le 10^3 $ random realizations
of static imperfections. In this way the total number of Floquet 
quasi-energies is $ N_D \times 2^{n_q} \approx 10^{4}$, 
which is large enough to 
get stable results in the study of spectral statistics.

\section{The ergodic regime}

\begin{figure} 
\centerline{
\begin{picture}(50,205)(0,0)
\put(-100,-5) {\epsfxsize=8.5cm \epsffile{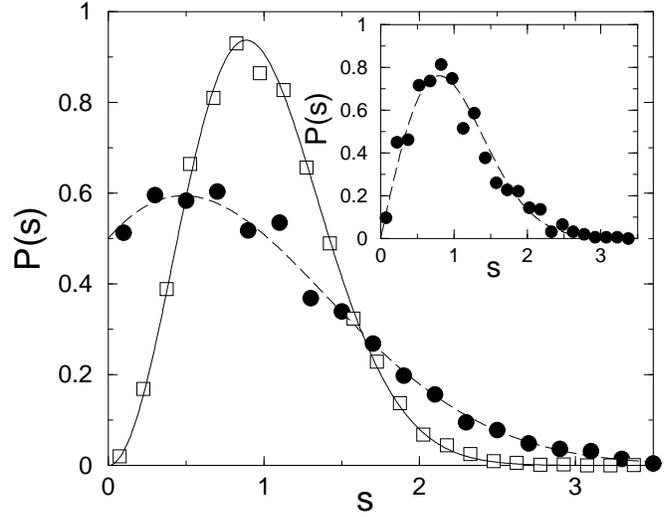}}
\end{picture}}
\caption{
The level spacing distribution for $K=\sqrt{2}$, $n_q=11$ and $J=0$, at 
$\epsilon =0$ (circles) and $\epsilon= 10^{-3}$ (squares).
The dashed and full lines give the theoretical distribution 
(\ref{2GOE}) and (\ref{WD}), respectively. 
Inset: statistics for one symmetry class at $\epsilon=0$ and same parameter 
values as in the main figure. The dashed line is the theoretical distribution 
(\ref{GOE}).}
\label{distKrad2}
\end{figure}

We focus our attention first on the ergodic regime, where the eigenfunctions
of the unperturbed Floquet operator are given by a complex superposition of 
order $N=2^{n_q}$ 
quantum register states. In particular, we consider $K=\sqrt{2}$, where the 
corresponding classical sawtooth map (\ref{clmap}) is completely chaotic.
In the ideal case in the absence of static imperfections ($\epsilon=0$, 
$\rho=0$),
the eigenvalues of the Floquet operator are divided in two different 
symmetry classes, even and odd with respect to the transformation 
$ n \rightarrow -n $. Thus the Floquet operator has two blocks which 
can be studied independently.
A convenient tool to characterize the spectral properties 
of the system is the level spacing statistics $P(s)$, giving the probability
to find two consecutive eigenvalues whose energy difference, 
normalized to the average level spacing, is in $[s,s+ds]$.
Since we are in the ergodic regime, the level spacing statistics
for each block is well described by the random matrix 
theory \cite{bohigas,guhr} in the presence of time-reversal symmetry: 
\begin{equation}
P_{O}(s)=\frac{\pi}{2} s e^{-\pi s^2/4}.
\label{GOE}
\end{equation}
This theoretical distribution is in agreement with the numerical results     
shown in the inset of Fig. \ref{distKrad2} for the symmetric class. 
The global spectral statistics can be computed from the single spectral 
statistics of each symmetry class \cite{bohigas}, and is given by:
\begin{equation}
P_{O}^{(2)}(s) = \frac{1}{2} \left( \hbox{erfc}\left(\frac{\sqrt{\pi}s}{4}
\right) 
\frac{\pi s}{4} e^{ - \pi s^2/16} + e^{ - \pi s^2/8} 
\right),
\label{2GOE}
\end{equation}
again in good agreement with the numerical data of Fig. \ref{distKrad2}.

\begin{figure} 
\centerline{
\begin{picture}(50,195)(0,0)
\put(-100,-5) {\epsfxsize=8.5cm \epsffile{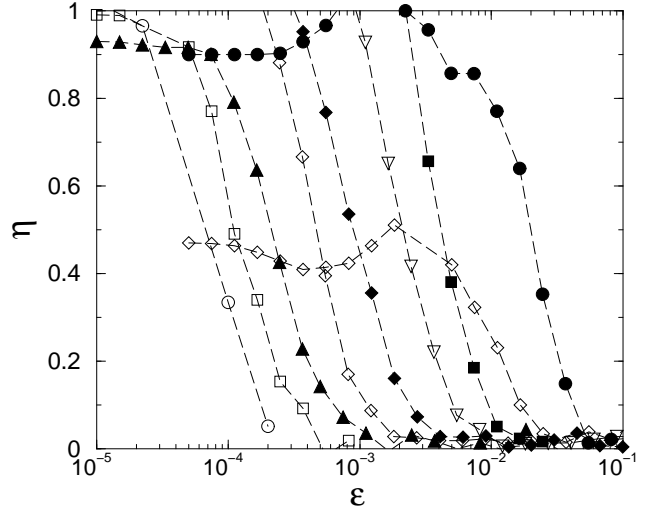}}
\end{picture}}
\caption{Dependence of the $\eta$ parameter on the scale imperfection strength
$\epsilon$ at $K=\sqrt{2}$, $J=0$. From right to left: $n_q=4,5,...,12$.}
\label{dst}
\end{figure}

Since static imperfections break the time reversal symmetry, the two 
symmetry classes become mixed and the system undergoes a crossover from 
(\ref{2GOE}) to a different random matrix statistics \cite{bohigas,guhr}:
\begin{equation} 
P_{U}(s)=\frac{32 s^2}{\pi^2} e^{- 4 s^2/\pi}.
\label{WD}
\end{equation}
The good agreement between this limiting distribution and the actual 
statistics computed from numerical data is shown in Fig. \ref{distKrad2} for
$n_q =11 $ and $\epsilon = 10^{-3}, J=0 $.
A convenient quantity to characterize the crossover from one 
limiting distribution to another is \cite{jacquod}
\begin{equation} 
\eta=\frac{\int_0^{s_0} (P(s)-P_{U}(s)) ds}{ \int_0^{s_0}
(P_{O}^{(2)}(s)-P_{U}(s)) ds}.
\end{equation}
where  $s_0=0.50285...$ is the first intersection point of 
$P_{O}^{(2)}(s)$ and
$P_{U}(s)$. This parameter goes from one to zero, when the spacing
probability changes from (\ref{2GOE}) to (\ref{WD}).
The behavior of $\eta$ as a function of $\epsilon$ is shown in Fig. \ref{dst},
for different $n_q$ values. This figure shows that it drops to zero when the 
imperfection strength grows. For  $\epsilon \to 0$ and small $n_q$, there are 
significant deviations from the value $\eta=1$ corresponding to the 
limiting distribution (\ref{2GOE}). This is due to the fact that the  
$\epsilon=0$ statistics includes only $2^{n_q}$ levels spacings. 
For very small $\epsilon$ values the statistics remains poor since 
each static imperfections realization gives essentially the same $P(s)$ 
distribution. Nevertheless, it is clear that $\eta(\epsilon=0)$ goes to one
when the number of qubits increases.

\begin{figure} 
\centerline{
\begin{picture}(50,195)(0,0)
\put(-100,-3) {\epsfxsize=8.5cm \epsffile{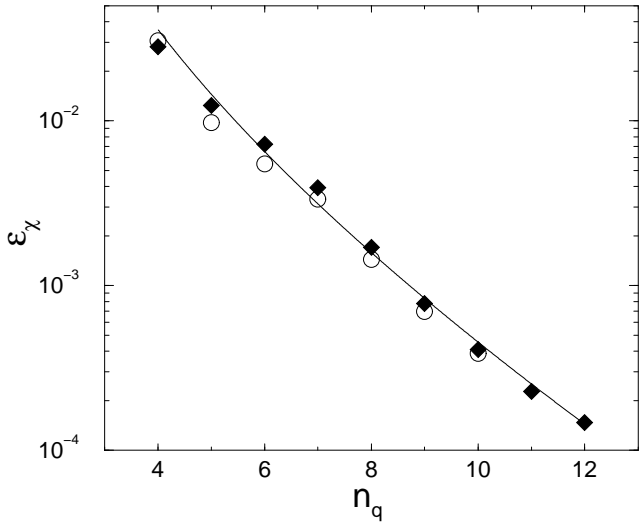}}
\end{picture}}
\caption{
Dependence of the imperfections strength $\epsilon_\chi$ at which $\eta=0.2$ 
on the number of qubits,  for $K= \sqrt{2}$, $J=0$ (diamonds) and $J=\delta$ 
(circles). The line gives the theoretical dependence $\epsilon_\chi= A  
2^{-n_q/2} n_q^{-2.5}$, with the fitting constant $A=4.3$. }
\label{eta}
\end{figure}

In order to study the dependence of the 
$\eta$-crossover on the number of qubits, we define 
the critical imperfections strength $\epsilon_\chi$ as the $\epsilon$ 
value at which  $\eta=0.2$ (similar results are obtained for different
$\eta$ values). In Fig. \ref{eta} we show that $\epsilon_\chi$ drops 
exponentially with the number of qubits. This exponential threshold is due
to the fact that, in the ergodic regime, quantum eigenstates are given
by a complex superposition of an exponentially large number of quantum 
register states. Indeed, due to quantum chaos, the eigenstates of the
unperturbed ($\epsilon=0$, J=0) Floquet operator 
(\ref{floquet}) can be written as 
\begin{equation}
\phi_\alpha^{(0)}= \sum_{m=1}^N c_\alpha^{(m)} u_m.
\end{equation}
Here $u_m$ are the quantum register states, and $c_\alpha^{(m)}$ are 
randomly fluctuating components, with $|c_\alpha^{(m)}| \sim 1/\sqrt{N}$ due
to wave function normalization.
The transition matrix elements between unperturbed eigenstates 
can be computed in perturbation theory. For $J=0$, they have 
a typical value 
\begin{eqnarray}
V_{\rm typ} \sim |\langle\phi_\beta^{(0)}|
\sum_{i=1}^{n_q}\delta_i\hat{\sigma}_i^z \tau_g n_g
| \phi_\alpha^{(0)}\rangle| \nonumber 
\\ \sim  
\tau_g n_q^{2} |\sum_{m,n=1}^N c_\alpha^{(m)} c_\beta^{(n)\star}
\sum_{i=1}^{n_q} \delta_i 
\langle u_n | \hat{\sigma}_i^z | u_m \rangle | 
\label{eqlunga}\\
\sim \epsilon n_q^{5/2} |\sum_{m=1}^N c_\alpha^{(m)} c_\beta^{(m)\star}|
\sim \epsilon n_q^{5/2} {N^{-1/2}}. \nonumber
\end{eqnarray}
In this expression, the typical phase error is $\delta \sqrt{n_q}$ (sum of 
$n_q$ random detunings $\delta_i$'s) and $\tau_g n_g \sim \tau_g n_q^2$ is the 
time used by the quantum computer to simulate one map step.
The last estimate in (\ref{eqlunga}) results from the sum of $N$ terms of
amplitude $|c_\alpha^{(m)} c_\beta^{(m)\star}| \sim 1/N$ and random phases.
Since the spacing between quasi-energy eigenstates is
$\Delta E \sim 1/N$, the threshold for the breaking of
perturbation theory can be estimated as
\begin{equation}
V_{\rm typ}/\Delta E \sim \epsilon_\chi n_q^{5/2}\sqrt{N} \sim 1.
\end{equation} 
The analytical result 
\begin{equation} 
\epsilon_\chi\sim \frac{1}{n_q^{5/2}\sqrt{N}}
\label{ergest} 
\end{equation} 
is confirmed by the numerical data of Fig. \ref{eta}.
For the case $J=\delta$, the threshold $\epsilon_\chi$ 
approximately decreases by a factor
$1.2$ with respect to the $J=0$ case (see again Fig. \ref{eta}), 
since residual inter-qubits interactions 
introduce further couplings between Floquet eigenstates. However, the same
functional dependence (\ref{ergest}) takes place.

We note that the same exponential sensitivity to static imperfections was
detected for the Floquet eigenstates in Ref. \cite{simone2}. This
confirms that spectral statistics are a useful tool to characterize the 
mixing of unperturbed eigenstates in an operating quantum computer.

\section{The quasi-integrable regime}

We now focus our study on the regime of quasi-integrability of the
sawtooth map, in which the system is not chaotic (i.e. the Lyapunov exponent 
is zero) and there is a non integrable component in the phase space.
In this situation, the system exhibits interesting physical properties, 
such as anomalous diffusion and hierarchical phase 
space structures \cite{simone1}. In this Section, we consider the case 
$K=-0.1$ and we study the transition to chaos for the eigenstates
of a quantum computer simulating the quantum sawtooth map. 

\begin{figure} 
\centerline{
\begin{picture}(50,190)(0,0)
\put(-110,-15) {\epsfxsize=9.2cm \epsffile{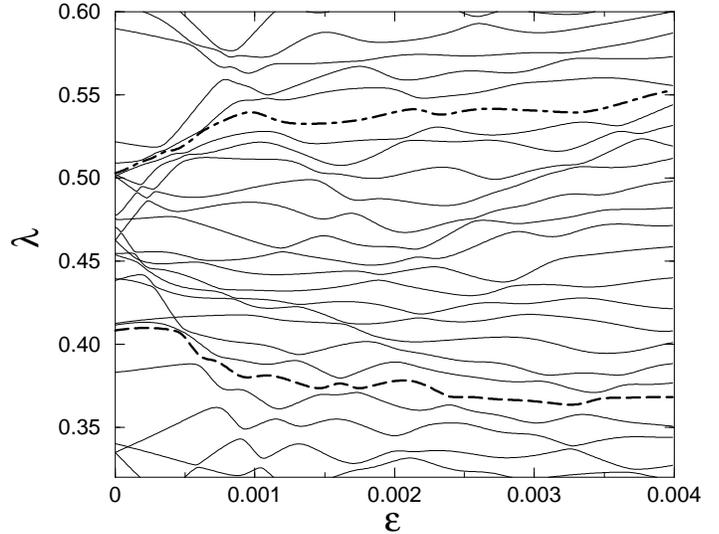}}
\end{picture}}
\caption{Dependence of quasi-energy eigenvalues on the imperfections strength
$\epsilon$, for $n_q=9$, $K=-0.1$, $J=0$. The  thick dashed 
(dot-dashed) curves corresponds to the eigenvector 
represented in Fig. \ref{husimi} right (left).}
\label{spag}
\end{figure}

\begin{figure} 
\centerline{\epsfxsize=4.cm \epsffile{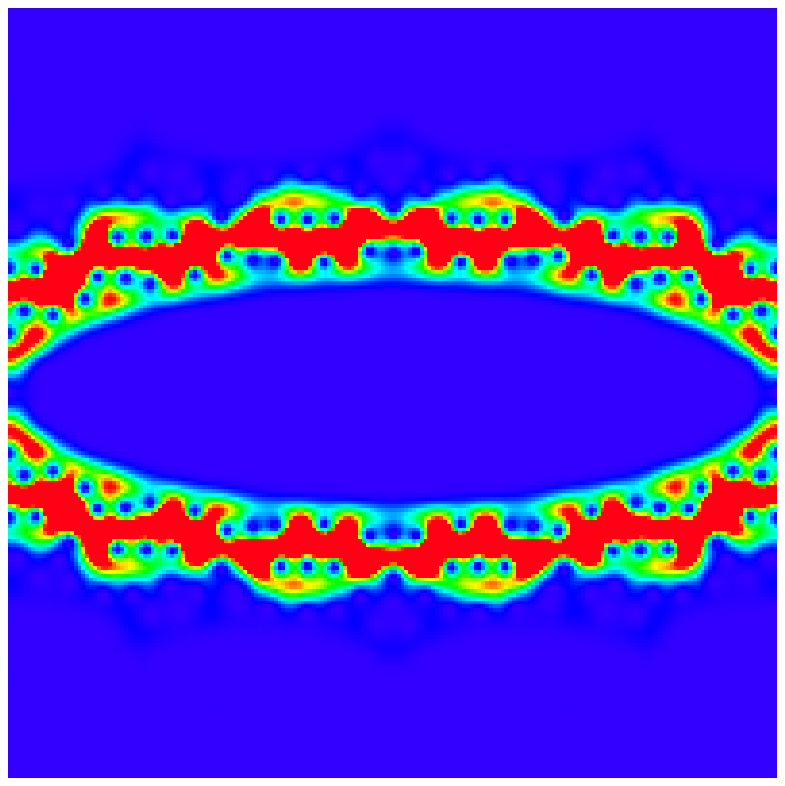}
\epsfxsize=4.cm\epsffile{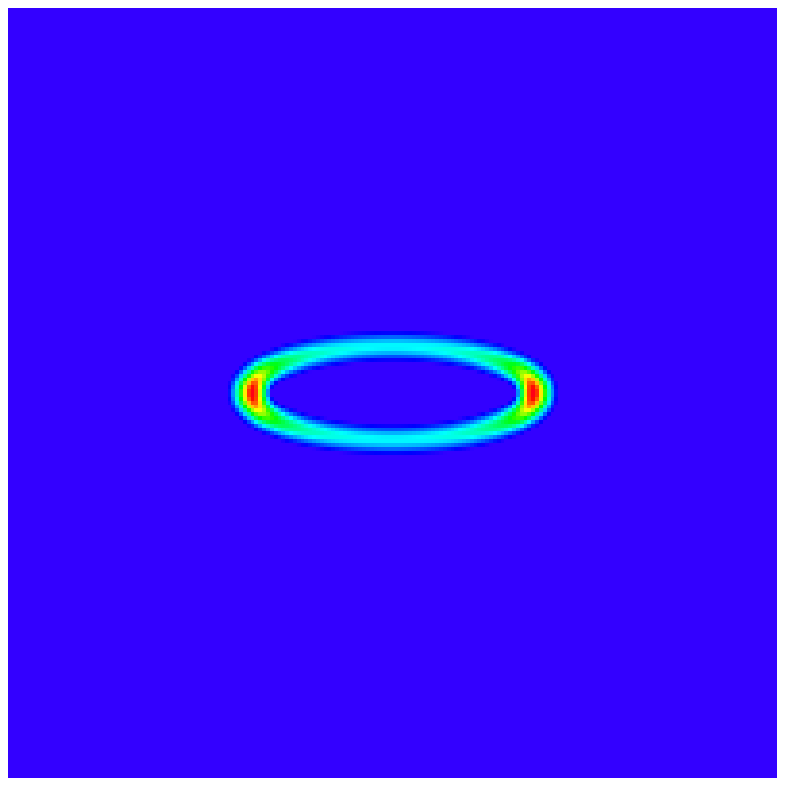}}
\centerline{\epsfxsize=4.cm \epsffile{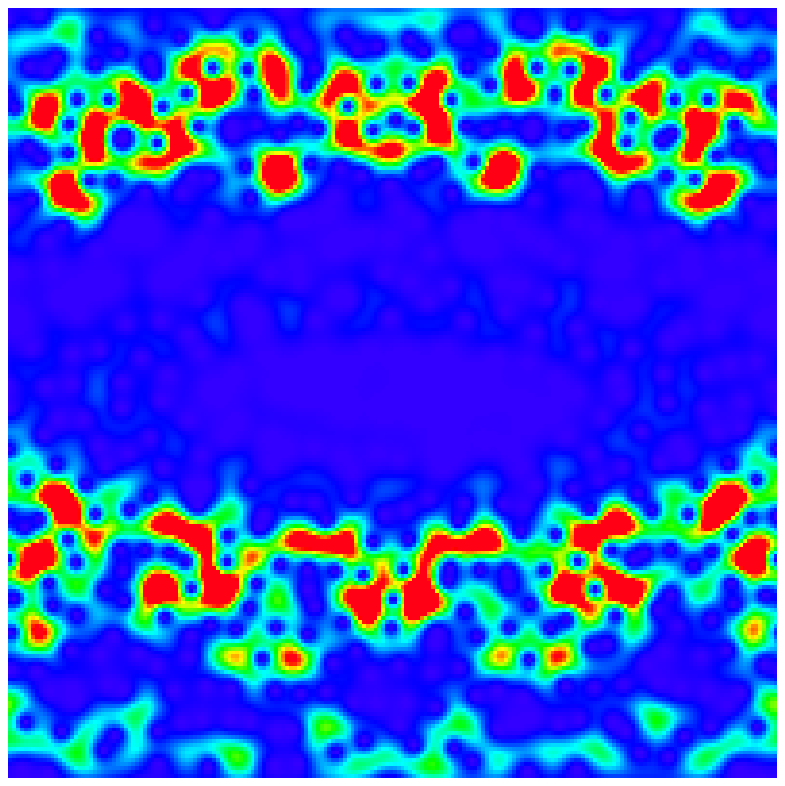}
\epsfxsize=4.cm\epsffile{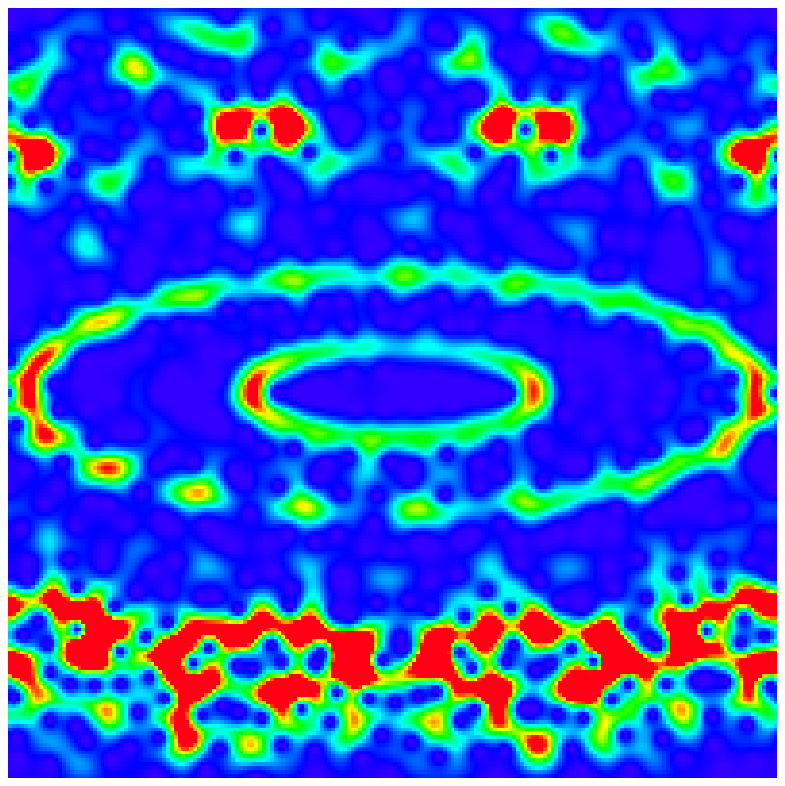}}
\centerline{\epsfxsize=4.cm \epsffile{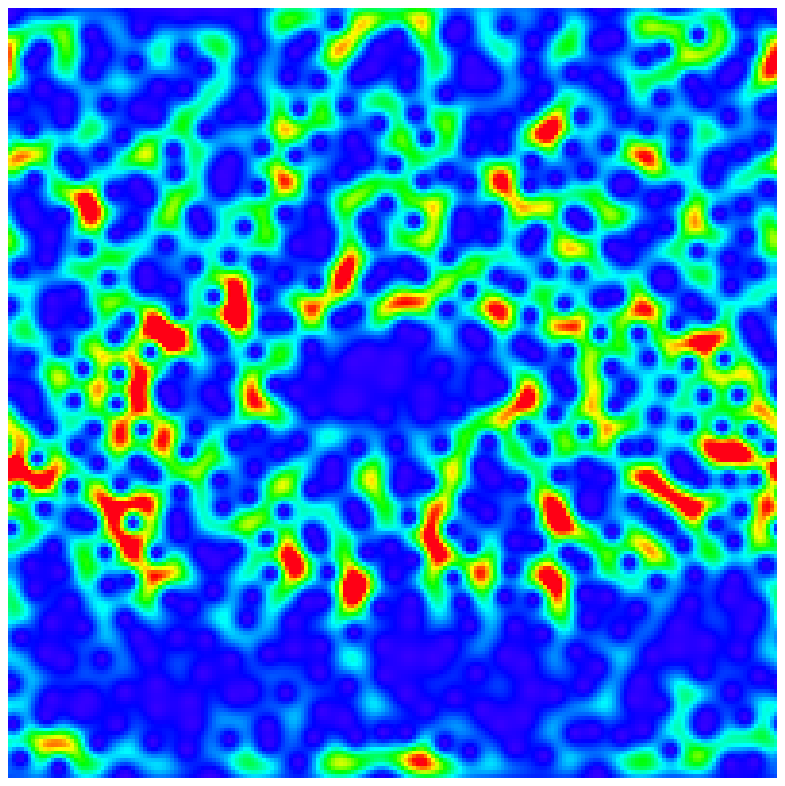}
\epsfxsize=4.cm\epsffile{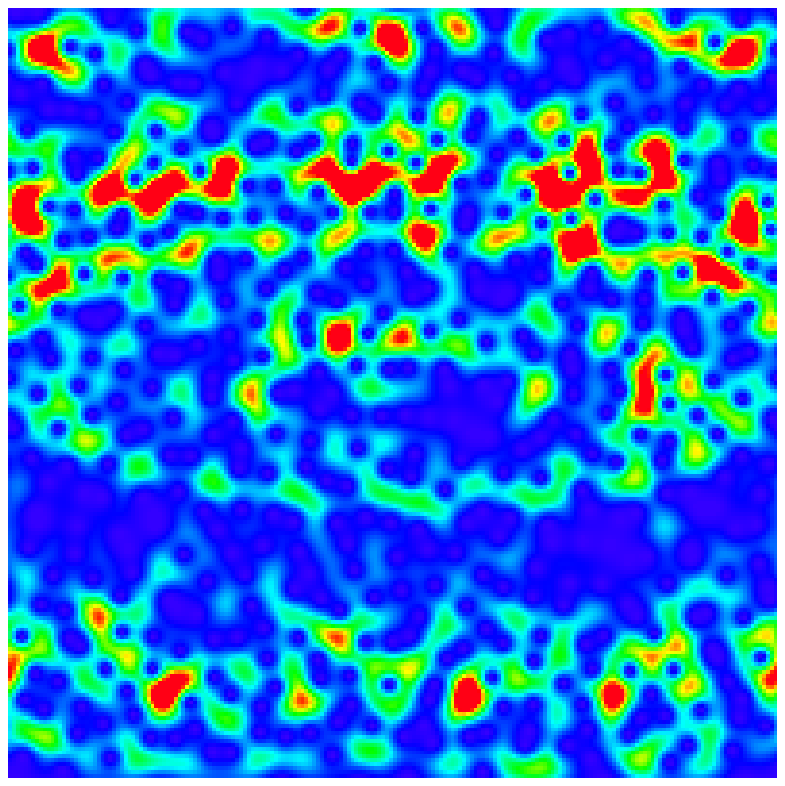}}
\caption{Husimi functions in action-angle variable $(p,\theta)$ ($-\pi \le p 
< \pi$ -vertical axis- and $0\leq \theta < 2\pi$ -horizontal axis-)
for the quantum sawtooth map at $n_q=9$, $K=-0.1$, $J=0$. 
The left (right) column corresponds to the dot-dashed (dashed) line in 
Fig. \ref{spag}, for  $\epsilon=0$ (top), $\epsilon =10^{-3}$ (middle),
 $\epsilon = 3 \times 10^{-3}$ (bottom).
We choose the ratio of the action-angle uncertainties
$s=\Delta p/\Delta\theta=T\Delta n/\Delta\theta=1$
($\Delta p \Delta\theta= T /2$).
The color is proportional to the density: blue for zero and
red for maximal density.
}
\label{husimi}
\end{figure}

A useful tool to demonstrate the transition to chaos induced by 
static imperfections is the parametric dependence of the quasi-energy
eigenvalues. The evolution of a part of the spectrum as a function of
the imperfection strength $\epsilon$ (at $J=0$) is shown in 
Fig. \ref{spag}. 
It makes evident the qualitative change induced by static imperfections.
At small $\epsilon$ the spectrum exhibits quasi-degeneracyes, while at large 
$\epsilon$ avoided crossings appear. Indeed, at small $\epsilon$, Floquet 
eigenstates with very close eigenvalues may lay so far apart that their 
overlap is negligible. Thus there is essentially no level repulsion for these 
eigenvalues. On the contrary, at large $\epsilon$, Floquet eigenstates are 
delocalized and therefore their overlap is significant, and  
induces level repulsion \cite{haake}. The delocalizing effect of static 
imperfections is evident in the Husimi functions of Fig. \ref{husimi}, 
drawn from two typical Floquet eigenstates \cite{husimi,husimi1}.

The two top plots of  Fig. \ref{husimi} represent two exact eigenstates
($\epsilon=0$, $J=0$). In the right picture, the quantum probability 
is concentrated
around an ellipse corresponding to a classical integrable trajectory (a torus 
in the phase space $(\theta,p)$). We note that the map (\ref{clmap}) is the 
discretized time evolution for an harmonic oscillator. Therefore it gives 
elliptic trajectories as far as border effects can be neglected. 
A completely different 
kind of eigenvector appears on the left picture. The Husimi function
is spread, and this reflects the properties of the corresponding classical
non-integrable trajectories which  diffuse in a non-Brownian way 
\cite{simone1}. We note that the number of such eigenstates is 
non-negligible. Indeed, 
we have checked that the probability of finding a diffusive classical 
trajectory starting from a randomly chosen initial condition is 0.12. 
We also note that classical diffusion is suppressed by quantum 
interference effects giving a distribution localized in momentum. 
The remaining pictures of Fig. \ref{husimi} represent the same Floquet 
eigenstates, computed in the presence of static imperfections.
The middle figures are obtained slightly below the chaos border  
induced by static imperfections,  
and maintain same similarities with the exact eigenstates. On the contrary, 
in the bottom pictures, taken above the chaos border, the eigenfunctions
are spread in all the available phase space and any structure has been
destroyed. We note that, starting from completely different unperturbed 
eigenfunctions, one gets statistically indistinguishable eigenfunctions 
with randomly fluctuating components. 

We now characterize the transition by studying the spectral statistics. 
In this regime, the limiting $P(s)$ distribution at 
$\epsilon=0$ is non trivial. 
Indeed, the Shnirelman theorem states that for a 
nearly integrable 
system the level spacing statistics $P(s)$ 
exhibits a huge peak near the origin ($s=0$) \cite{shni,chirdima}.
In the following we choose to eliminate such peak, related to time-reversal 
invariance ($n \to -n$, $\theta \to 2\pi-\theta$). 
To do that we operate the transformation \cite{felice}
\begin{eqnarray}
\left\{
\begin{array}{l}
 n  \to  n + \phi, \\
\cr
 \theta  \to \theta + \theta_0, 
\end{array}
\right. 
\end{eqnarray}
where $\phi$ plays the role of an Aharonov-Bohm flux \cite{note}. 
The spectral statistics is still complex since in the classical limit 
the phase space has two components, integrable and non integrable. 
In the semi-classical limit, the levels belonging to these two components  
become uncorrelated and the global spacing statistics is 
given by the superposition of each component's statistics \cite{berry}. 
We note that in this case both components have non-trivial statistics. 
In particular it has been shown that for the related harmonic oscillator 
case the quasi-energy spectral statistics exhibits level repulsion
and a finite number of peaks \cite{bohigas1}, instead of the Poisson 
statistics typical of integrable systems. 

\begin{figure} 
\centerline{
\begin{picture}(50,195)(0,0)
\put(-100,-5) {\epsfxsize=8.5cm \epsffile{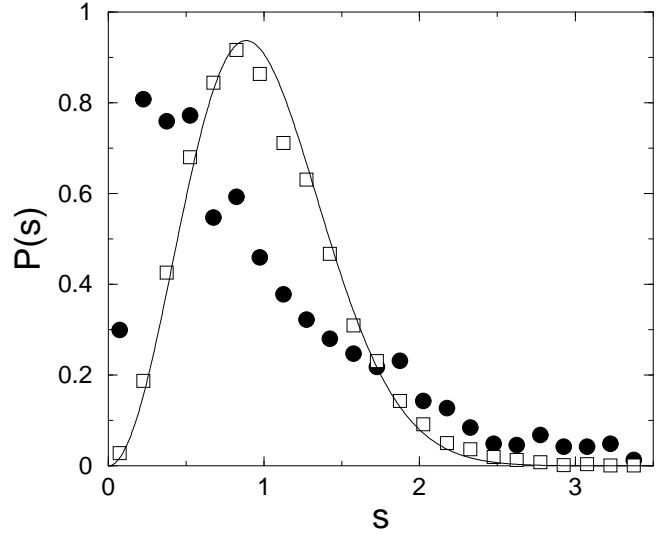}}
\end{picture}}
\caption{Level spacing distribution for $n_q=11$, $K=-0.1$,
$\theta_0=\phi=\sqrt{2}/5$, $J=0$, $\epsilon=0$ (circles) 
and $\epsilon=10^{-3}$ (squares).
The solid line gives the Wigner-Dyson distribution (Eq. (\ref{WD})).}  
\label{distK-01}
\end{figure}

\begin{figure} 
\centerline{
\begin{picture}(50,195)(0,0)
\put(-100,-5) {\epsfxsize=8.5cm \epsffile{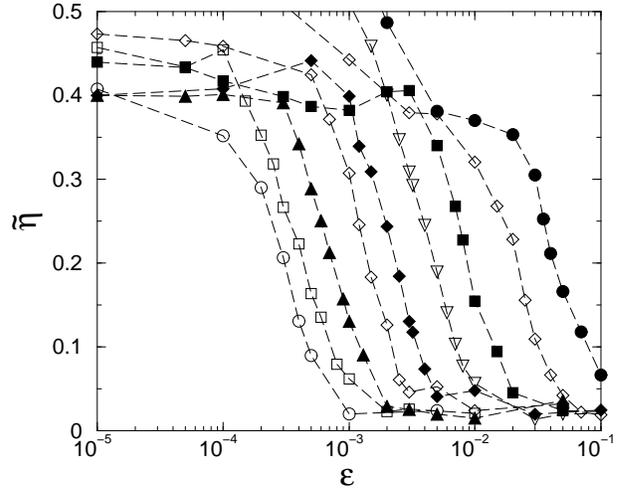}}
\end{picture}}
\caption{Dependence of the parameter $\tilde \eta$ on the imperfection strength
$\epsilon$ for  $K=-0.1$, $\theta_0=\phi=\sqrt{2}/5$, $J=0$.
From right to left: $n_q=4,5,\dots,12$.}
\label{dstK-01}
\end{figure}

The spectral statistics at $\epsilon=0$, $J=0$ is shown in  
Fig. \ref{distK-01}, 
for $n_q=11$. In the same figure, one can see that static imperfections 
induce a crossover to the Wigner-Dyson distribution (\ref{WD}). One can 
compare this figure with Fig. \ref{distKrad2}. The $\epsilon=0$ 
distributions are completely different, reflecting different dynamical
regimes. On the contrary, the same universal distribution is found in
the regime in which static imperfections destroy all symmetries. 
Even though in this case we cannot provide an analytical 
expression for the $\epsilon=J=0$ statistics, 
the crossover to the Wigner-Dyson distribution (\ref{WD}) 
can be characterized by the parameter
\begin{equation}
\tilde \eta =\left(\int_0^{+\infty} [ P(s) - P_{U}(s) ]^2  ds
\right)^{1/2}, 
\label{etatilde}
\end{equation}
measuring the distance of the spacing distribution from (\ref{WD}).
The behavior of $\tilde \eta$ as a function of $\epsilon$ (at $J=0$) 
is shown in Fig. \ref{dstK-01} for various  $n_q$ values.
This figure shows again that the Wigner -Dyson distribution 
($\tilde \eta=0$) is reached by increasing $\epsilon$.

\begin{figure} 
\centerline{
\begin{picture}(50,195)(0,0)
\put(-100,-5) {\epsfxsize=8.5cm \epsffile{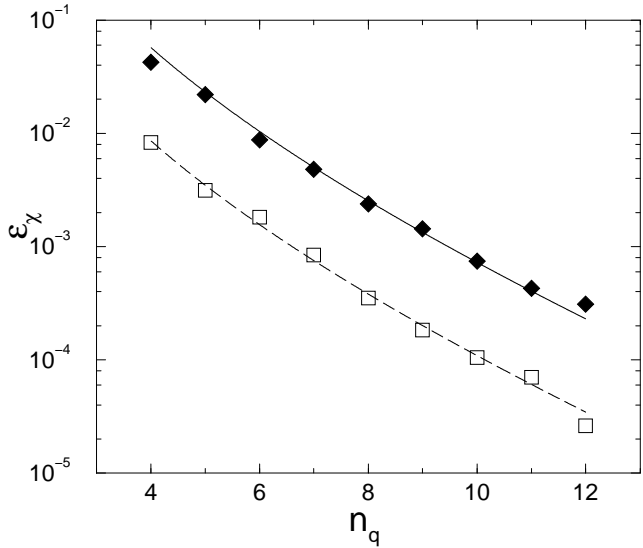}}
\end{picture}}
\caption{Dependence of the imperfections strength $\epsilon_\chi$ at
which $\tilde \eta =0.2$ (see Fig. \ref{dstK-01}) on the number of qubits
$n_q$ (diamonds), for  $K=-0.1$, $\theta_0=\phi=\sqrt{2}/5$, $J=0$. 
Squares give the critical imperfection strengths at which the quantum 
eigenstates entropy $S=1$. Lines
give the theoretical dependency $\epsilon_\chi = B  N^{-1/2}
n_q^{-5/2}$, with the fitting constant $B=1$ (below) and 
$B=6.5$ (above).}
\label{epschi2}
\end{figure}

Similarly to what we have done in the ergodic regime, we define a 
critical imperfections strength $\epsilon_\chi$ as the $\epsilon$ value
at which $\tilde \eta=0.2$.
The dependence of $\epsilon_\chi$ on the number of qubits is shown in Fig.
\ref{epschi2}. Quite surprisingly, even in this quasi-integrable regime we
can fit the exponential drop of $\epsilon_\chi$ with $n_q$ via the 
analytical function $\epsilon_\chi = B 2^{-n_q/2} n_q^{-5/2}$, with the 
fitting constant $B=6.5$.
Therefore, it seems that the theoretical argument developed in the previous
Section for ergodic eigenstates has a broader validity and can apply also
in the more general mixed phase space dynamics. 

The same exponential 
sensitivity to static imperfections can be detected also in the Floquet 
eigenstates. The mixing of unperturbed eigenstates, induced by static 
imperfections, is characterized by the quantum eigenstates entropy
\begin{equation} 
S_\alpha=-\sum_{\beta=1}^N p_{\alpha\beta}\log_2 
p_{\alpha\beta},  
\end{equation} 
where $p_{\alpha\beta}=|\langle\phi_\beta^{(0)}|
\phi_\alpha^{(\epsilon,\rho)}\rangle|^2$.
In this way $S_\alpha=0$ 
if $\phi_\alpha^{(\epsilon,\rho)}$ coincides with one eigenstate
at $\epsilon=\rho=0$, $S_\alpha=1$ if $\phi_\alpha^{(\epsilon,\rho)}$
is equally composed of two ideal ($\epsilon=\rho=0$) eigenstates, and 
$S_\alpha=n_q$ (maximal value) if all
$\phi_\beta^{(0)}$ $(\beta=1,...,N=2^{n_q})$ contribute
equally to $\phi_\alpha^{(\epsilon,\rho)}$. In Fig. \ref{epschi2} we show that 
the critical imperfection strength at which $S=1$ drops 
exponentially with the number of qubits ($S$ is the average of $S_\alpha$ 
over $\alpha$ and static imperfection realizations). This shows that 
the transition to chaos can be characterized both by the mixing of 
unperturbed eigenstates and by the transition in the spectral statistics. 

\section{The integrable regime}

We now consider the integrable case $K=-1$. Independently 
of $n_q$, the quasi-energy 
spectrum is composed of 6 degenerate levels (see Fig.9),  
$\lambda^{(j)}=(2\pi/6)j$
(up to an unessential global phase factor).
This phenomenon can be explained from classical mechanics:
indeed the 6-th iterate of the map (\ref{clmap}) gives 
the identity, i.e. all orbits are periodic with period 
at most 6 \cite{creutz}. 
The same phenomenon can be observed at $K=-2$ 
($K=-3$), where the spectrum has a degeneracy 4 (3) and 
the 4-th (third) iterate of (\ref{clmap}) is the identity.   
In these cases, the quantum evolution and the classical 
discretized evolution {\it coincide}: 
this means that the quantum unitary evolution (\ref{qumap}) 
maps a given phase space distribution in the same way as the 
discretized Liouville operator does in classical mechanics   
(see, e.g., Ref. \cite{saraceno}). 

\begin{figure} 
\centerline{
\begin{picture}(50,195)(0,0)
\put(-100,-5) {\epsfxsize=8.5cm \epsffile{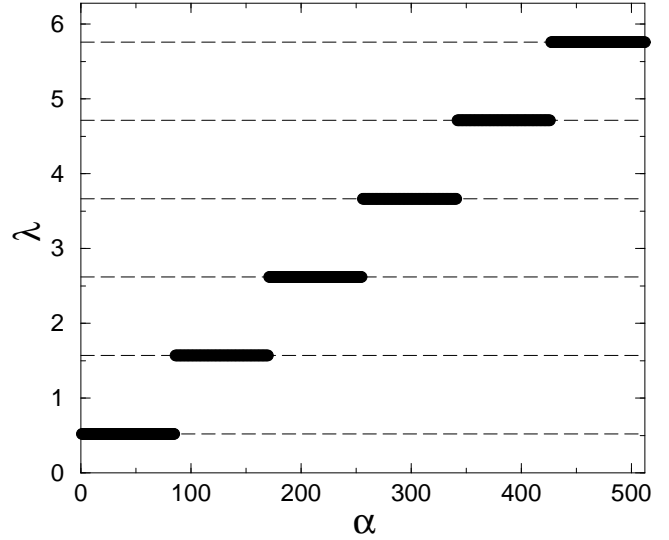}}
\end{picture}}
\caption{Quasi-energy levels $\lambda_\alpha$ at $n_q=9$, $K=-1$,
$\epsilon=J=0$ (thick bands) and theoretical 
values (dashed lines).} 
\label{levels}
\end{figure}

\begin{figure} 
\centerline{
\begin{picture}(50,195)(0,0)
\put(-100,-5) {\epsfxsize=8.5cm \epsffile{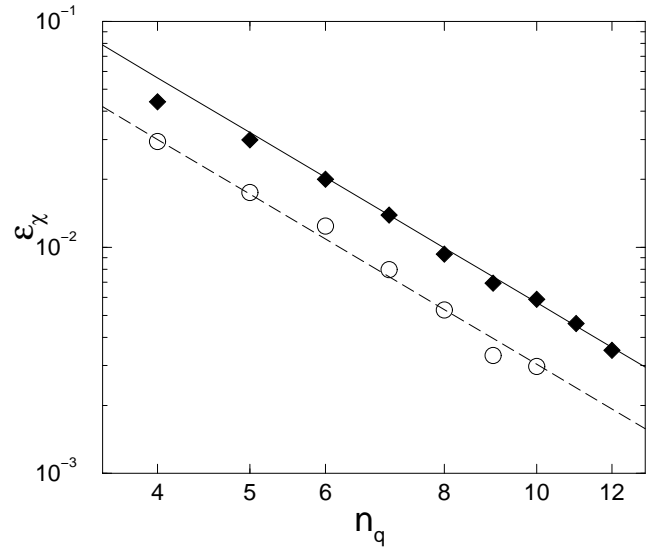}}
\end{picture}}
\caption{Dependence of the imperfections strength $\epsilon_\chi$ at
which $\tilde \eta =0.2$ on the number of qubits
$n_q$, for $K=-1$, $J=0$ (diamonds) and $J=\delta$ (circles);
note the  log-log scale.
Lines give the theoretical dependence $\epsilon_\chi = C 
n_q^{-5/2}$, with the fitting constant $C=1$ (below) and 
$C=1.8$ (above).}
\end{figure}

Static imperfections induce a crossover to the 
universal Wigner-Dyson distribution (\ref{WD}), 
that can be characterized by the parameter 
$\tilde \eta$ defined in Eq. (\ref{etatilde}). 
Again, one can define a critical imperfection strength 
$\epsilon_\chi$ as the value of $\epsilon$ at 
which $\tilde\eta (\epsilon_\chi)=0.2$. The result is 
shown in Fig. 10, where it is seen that $\epsilon_\chi$ 
drops polynomially with $n_q$: 
$\epsilon_\chi\propto n_q^{-5/2}$. 
This happens both for $J=0$ and $J=\delta$. 
We note that this algebraic dependence contrasts 
the exponential decay observed in previous Sections 
for the ergodic and quasi-integrable regimes. 
This can be understood via the following argument. 
Quantum ergodicity can be reached only when 
levels of the different bands of Fig. 9 are mixed. 
Non-degenerate levels are separated by an energy 
spacing $\Delta E$ which is $N$-independent. 
On the other hand, also the typical overlap 
between nearby eigenstates (in space) 
is $N$-independent (their space  
separation drops with $N$, but also their typical width 
drops with $N$).  
Since the typical error is $\delta\sqrt{n_q}$ and 
the time needed to simulate one map step is 
$\sim \tau_g n_q^2$ (see the discussion following 
Eq. (\ref{eqlunga})), one can estimate 
the typical transition matrix element 
$V_{\rm typ}\sim \epsilon n_q^{5/2}$, and the threshold 
for the breaking of perturbation theory as
\begin{equation}  
V_{\rm typ}/\Delta E \sim \epsilon_\chi n_q^{5/2} \sim 1, 
\end{equation} 
giving the analytical estimate 
\begin{equation} 
\epsilon_\chi\sim n_q^{-5/2}. 
\end{equation} 

\section{Conclusions}

In this paper, we have studied the transition to quantum 
chaos, induced by static imperfections, in the spectral 
statistics of an operating quantum computer.
The threshold for the transition to chaos drops 
exponentially with the number of qubits, both in 
the ergodic and in the more general quasi-integrable 
regime. On the contrary, in the integrable regime 
the chaos border drops only 
algebraically with the number of qubits.  
This is due to the presence of a finite number of bands 
in the spectrum of the time evolution operator
related to global periodicity of classical dynamics. 
We note that similar strong spectral degeneracyes 
have been observed in the Grover's algorithm and 
in the quantum Fourier transform in Ref. \cite{braun}.  
We also stress that a complex dynamical system is 
generically quasi-integrable. The simulation 
of such systems is well accessible to the first generation
of quantum computers with less than 10 qubits and 
we think that this class of 
quantum algorithms deserves further studies. 

This work was supported in part by the EC RTN contract
HPRN-CT-2000-0156 
and by the NSA and ARDA under
ARO contracts No. DAAD19-01-1-0553 (for D.L.S.)  
and No. DAAD19-02-1-0086. 
Support from the 
PRIN-2000 ``Chaos and localization in classical and
quantum mechanics'' is gratefully acknowledged.


\begin{thebibliography}{99}
\bibitem{husimi1} In order to stress the perturbation induced 
symmetry breaking, in Fig. \ref{husimi} we consider the 
time-symmetric version of the map (\ref{qumap}):   
\begin{equation}
\overline{\psi}=\hat{U}\psi =
e^{-iT\hat{n}^2/4}
e^{ik(\hat{\theta}-\pi)^2/2}
e^{-iT\hat{n}^2/4} \psi.   
\end{equation}
This map has axial symmetries, $\theta\to 2\pi-\theta$ or $p\to -p$, 
while the evolution operator (\ref{qumap}) has only central symmetry,  
$\theta\to 2\pi-\theta$ and $p\to -p$.
\bibitem{note} This transformation does not change the number of elementary 
gates per map iteration.
\end{thebibliography}
\end{document}